\documentstyle[11pt,MyM,epsfig]{article}

\def\Journal#1#2#3#4{{#1} {\bf #2}, #3 (#4)}

\def\NPB{{\em Nucl. Phys.} B}
\def\PLB{{\em Phys. Lett.}  B}
\def\PRL{\em Phys. Rev. Lett.}
\def\PRD{{\em Phys. Rev.} D}

\def\ra{\rightarrow}

\def\be{\begin{equation}}
\def\ee{\end{equation}}
\def\bea{\begin{eqnarray}}
\def\eea{\end{eqnarray}}

\def\ra{\longrightarrow}

\def\eq{\begin{equation}}
\def\eqx{\end{equation}}
\def\eqn{\begin{eqnarray}}
\def\eqnx{\end{eqnarray}}

\newcommand{\al}{\alpha}
\newcommand{\bt}{\beta}
\newcommand{\gm}{\gamma}
\newcommand{\qq}{\overline{q}}
\newcommand{\ppsi}{\Psi(z,\zb)}
\newcommand{\zb}{\overline{z}}
\newcommand{\lm}{\lambda}
\newcommand{\QQ}{\tilde{Q}}
\newcommand{\uu}{\overline{u}}
\newcommand{\f}[2]{{#1}/{#2}}
\newcommand{\gms}{\gamma^*}

\begin{document}
\vspace*{4cm}

\title{The perturbative odderon intercept\footnote{Presented by R.A. Janik
at the 1999 Rencontres de Moriond} }

\author{\underline{R.A. Janik${}^{a,b}$}, J. Wosiek${}^b$}

\address{${}^a$ Service de Physique Th{\`e}orique, CEA Saclay\\
91191 Gif sur Yvette, France,\\
${}^b$ Institute of Physics, 
Jagellonian University,\\ Reymonta 4,30-059 Cracow, Poland\\
E-mail: janik@spht.saclay.cea.fr, wosiek@thrisc.if.uj.edu.pl} 


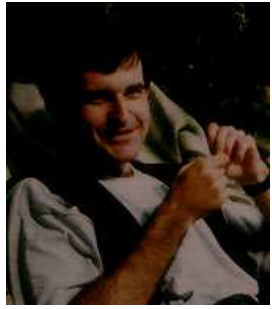
\maketitle\abstracts{We present our recent results on the odderon
intercept in perturbative QCD, obtained through the solution of the Baxter
equation and investigation of the spectrum of the relevant constant of
motion.
}

\section{Introduction}

An interesting problem of QCD is to understand the behaviour of the
theory in the Regge limit of large energy, fixed momentum transfer. In
the Leading Logarithmic Approximation the leading pole in the $C=+1$
channel is the famous BFKL pomeron \cite{BFKL}. Later this was
generalized to the channel odd under charge conjugation ($C=-1$) ---
the odderon \cite{odd}. In contrast to the BFKL case, however,
the value of the intercept remained unknown despite the discovery of
conformal symmetry and  integrals of motion \cite{lipq}.

The hard odderon may be observed in such processes as $\gms
\gms\ra\eta_c+X$, $\gms p\ra\eta_c+X$, $\gms\gms\ra \eta_c\eta_c$.
A number of recent papers \cite{phen} calculated the cross sections
in the approximation where the odderon was modelled by an exchange of three
(non-reggeized) gluons. The cross sections were small, and hence it is
interesting to see if reggeization could lead to a significant
enhancement at high energies, which could perhaps be investigated
experimentally. 

Recently substantial progress \cite{lip,fadkor} has been made with the
reduction of the problem to solving the Baxter equation
\eq
\label{e.baxter}
(\lm+i)^3 Q(\lm+i)+(\lm-i)^3 Q(\lm-i)=(2\lm^3+q_2\lm+q_3)Q(\lm)
\eqx
for physical values of the constants of motion $q_2$ and $q_3$. 
The intercept then follows through a logarithmic derivative \cite{fadkor}.
%
A number of approximation techniques for solving this equation have been
tried \cite{fadkor,KorQ}. In our work \cite{usI} an exact method of solving the
Baxter equation for arbitrary values of the parameters $q_2$ and $q_3$
was developed. This reduced the problem to finding the physical
spectrum of $q_3$ ($q_2$ is fixed by group theory to be $q_2=h(1-h)$
where $h$ is the conformal weight).  

Ideally one could do this directly, by requiring that the solution of
the Baxter equation $Q(\lm)$, gives rise to a normalizable, single-valued
wavefunction (see below).   
However, the explicit relation between wavefunctions and the
functional forms of the solutions $Q(\lm)$ is very indirect
\cite{fadkor}. In particular it is still unknown 
how does the
normalizability requirement explicitly look like in terms of $Q(\lm)$.

In our second paper \cite{usII} we pursued a more direct approach,
using standard physical requirements which are imposed on the wave
function to find the allowed values of $q_3$ and the explicit form of
the wave function. This form can be useful when considering the
coupling of the odderon to external probes.
In
this talk we would like to present the main results of these papers
and give our conclusions on the intercept of the odderon.



\section{Solution of the Baxter equation}

The Baxter equation poses a number of
difficulties. It is a nonlocal functional equation which possesses
polynomial solutions only for integer values of the conformal weight
$h=n>3$. In contrast to the BFKL case there is no easy analytical
continuation to the physically most interesting case of $h=1/2$.  
Quasiclassical methods \cite{KorQ} involve an expansion
in $1/h$ and use powerful methods to continue to $h=1/2$, the
precision is, however, difficult to control. 
Our aim was to find an expression for the solution directly for
$h=1/2$ (or indeed also for more general $h$) in a form which could be
numerically calculated to an arbitrary precision.

The starting point was the contour integral representation used in
\cite{fadkor}:
\eq
\label{e.int}
Q(\lm)=\int_C \frac{dz}{2\pi i} z^{-i\lm-1}(z-1)^{i\lm-1} \QQ(z)
\eqx
where $\QQ(z)$ satisfies a $3^{rd}$ order differential equation $D\QQ(z)=0$.
The expression (\ref{e.int}) satifies then the Baxter equation only if the contour
$C$ is choosen as such that the boundary terms arising from
integration by parts cancel out. It turns out that for $h=1/2$ it is
impossible to choose such a contour because the curve always ends up
on a different sheet of the Riemann surface of the integrand. 
To remedy this we extended the
ansatz (\ref{e.int}) to a sum of two integrals \cite{Acta,usI}:
\eq
\label{e.twoint}
\int_{C_{I}}\frac{dz}{2\pi i}K(z,\lm) \QQ_1(z)+
\int_{C_{II}}\frac{dz}{2\pi i} K(z,\lm) \QQ_2(z)
\eqx
where the contours are independent (see {\em e.g.} figures in
\cite{Acta,usI}).   
The functions $\QQ_1(z)$ and $\QQ_2(z)$ are both solutions of
$D\QQ(z)=0$, and thus depend initially on 6
free parameters.
We now impose the condition of cancellation of boundary terms. This
gives 3 equations, leaving us with $6-3=3$ parameters. Now one notes
\cite{usI} that since $D\QQ(z)=0$ has a solution holomorphic
at infinity, it gets integrated out to zero in each of the integrals
in (\ref{e.twoint}). This leaves us with only $3-2=1$ parameter which
is just an irrelevant normalization. The above procedure leads
therefore to a unique solution within our ansatz (\ref{e.twoint}).
This solution can be calculated to an arbitrary precision by
including a sufficient number of terms in power series expansions of
the $\QQ_i(z)$'s.
This being done we will move on, in the next section, to consider the
problem of finding the physical values of $q_3$.

\section{Quantization of $q_3$}

The wavefunction $\ppsi$ can be decomposed into the following sum:
\eq
\label{e.decomp}
\ppsi=\sum_{i,j}  \uu_i(\zb) A^{(0)}_{ij} u_j(z),
\eqx
where $u_i(z)$ and $\uu_j(\zb)$ are eigenfunctions (analytic in the whole
complex plane apart from some cuts) of the $3^{rd}$
order ordinary differential operators $\hat{q}_3$ and
$\hat{\qq}_3$ with eigenvalues $q_3$ and $\qq_3$. 
We will later also 
consider the $(-q_3,-\qq_3)$ sector which is necessary for Bose
symmetry of the wavefunction. 

We impose the following obvious physical requirements for the
wavefunction:
(1) $\ppsi$ must be single-valued,
(2) $\ppsi$ must be normalizable and 
(3) $\ppsi$ must satisfy Bose symmetry.
The crucial assumption is the first one. It turns out that
normalizability follows automatically (apart from the unphysical case
of $q_3=0$), and Bose symmetry is easy to implement.

We must examine the requirement of single valuedness near the singular
points $z=0$ and $z=1$ ($z=\infty$ follows --- see discussion in \cite{usII}).
Using the asymptotic behaviour of the $u_i$'s near $z=0$ ($u_1(z)\sim
z^{1/3}$, $u_2(z)\sim z^{5/6}$, $u_3(z)\sim z^{5/6}\log z +z^{-1/3}$)
it is clear that the coefficient matrix $A^{(0)}$ of (\ref{e.decomp}) must
have the form
\eq
\label{e.form}
 A^{(0)}=\left(    \begin{array}{ccc}
              \alpha & 0     & 0           \\
              0      & \beta & \gamma      \\  
              0      & \gamma& 0           \\
              \end{array} \right).
\eqx
We thus have initially 3 parameters $\al$, $\bt$ and $\gm$. Now we
must impose the same requirement around $z=1$. To this end we note
that the solutions $v_i(z)\equiv u_i(1-1/z)$ have similar asymptotics
around $z=1$ to the asymptotics of $u_i$'s around $z=0$. One can
numerically calculate the analytical continuation matrices
$u_i(z)=\Gamma_{ij} v_j(z)$.

We must now reexpress the wavefunction $\ppsi$ in terms of the $v_i$'s
and require that the transformed coefficient matrix
$A^{(1)}=\overline{\Gamma} A^{(0)}\Gamma$ has the same form as
(\ref{e.form}). This leads to a  
number of linear homogeneous equations for $\al$, $\bt$ and
$\gm$. 
Because the number of equations is greater than 3, the
existence of a nonzero solution fixes both the parameters of the
wavefunction $\al$, $\bt$, $\gm$ {\em and} the allowed values of
$q_3$. If the coefficient matrices $A^{(0)}$ and $A^{(1)}$ coincide,
the requirement of Bose symmetry boils down to adding the
corresponding wavefunction in the $(-q_3,-\qq_3)$ sector \cite{usII}.
\eq
\ppsi=\Psi_{q_3,q_3^*}(z,\zb)+\Psi_{-q_3,-q_3^*}(z,\zb)
\eqx

\section{Results}

In \cite{usII} a number of possible solutions were found. The
requirement of Bose symmetry picked out just those lying on the
imaginary axis. We may now plug in those values of $q_3$ into our
solution of the Baxter equation to yield the intercepts corresponding
to those states. The results are summarized below:

  \begin{center}
   \begin{tabular}{|c|cc|cc|} \hline\hline
   {\em No.} & \multicolumn{2}{c|}{ $q_3$ } 
                                     & \multicolumn{2}{c|}{ $\epsilon_3$} \\
   \hline
  1 & \multicolumn{2}{c|}{$0.20526 i$}
                                     & \multicolumn{2}{c|}{$-0.49434$} \\
  2 & \multicolumn{2}{c|}{$2.34392 i$} 
                                    & \multicolumn{2}{c|}{$-5.16930$} \\
  3 & \multicolumn{2}{c|}{$8.32635 i$} & \multicolumn{2}{c|}{$-7.70234$}  \\ 
   \hline\hline  
   \end{tabular}
  \end{center}

We are thus led to conclude that the odderon state with the highest
intercept has $q_3=\pm 0.20526i$. The intercept of this state reads
\eq
\al_O=1-0.24717\f{\al_s N_c}{\pi}=1-0.16478\cdot\f{3\al_s N_c}{2\pi}
\eqx 
and the wavefunction parameters are $\al=0.7096$, $\bt=-0.6894$ and $\gm=0.1457$.
Recently this result has been confirmed by M.A. Braun \cite{braunlast} who
redid his earlier work on the variational functional using our wave function,
%
and obtained after a formidable numerical calculation 
$\al_O=1-0.16606\cdot\f{3\al_s N_c}{2\pi}$,
in excellent agreement with our result following from the Baxter
equation. A second check of our results follows from the new symmetry
of the odderon discovered by Lipatov \cite{Lipsym} which forces the
wavefunction parameters to be related by $\gm=|q_3|\al$ which indeed
is verified in our case.
A more precise check can be made using the very recent asymptotic
formula \cite{Lipsym} for the energy expressed in terms of $q_3$ 
derived by Lipatov
for large $q_3$. We cannot use it for the ground state but for the
higher states identified here one obtains $\epsilon_3=-5.16956$ and
$\epsilon_3=-7.70234$ (for states with $q_3=\pm 2.34392i$ and $q_3=\pm
8.32635i$ respectively) which is in perfect agreement with our earlier results
obtained using the Baxter equation.

The states with higher $q_3$ give subleading contributions as
expected. Also a change in $h$ from $h=1/2$ to $h=1/2+i\nu$ also
lowers the energy. This observation is consistent with the expectation
that the dominant contribution should come from the $h\leftrightarrow
1-h$ symmetric point $h=1/2$ investigated in this work.

Some further confirmations and results are
presented in \cite{braunlast}  and in \cite{Rostwor}.

As a final point we note that our results suggest that the predictions
\cite{phen} for odderon mediated processes will not be enhanced at
high energies. It remains to see if the specific form of the wave
function might lead to some interesting effects through the coupling
of the ground state of the odderon found here, to external probes.

\section*{Acknowledgments}
  This work is  supported by the Polish Committee for Scientific Research 
  under grant no. PB 2P03B08614 and PB 2P03B04412.

\end{document}